# Automatic Calculation of Resolution in Lateral Cephalogram Based on Scale Mark Detection

Jia Guo, Shumeng Wang, Huiqi Li*
School of Information and Electronics
Beijing Institute of Technology
Beijing, China

*Abstract*—Cephalometric analysis is an important tool for orthodontic diagnosis. At present, most cephalometric analysis is performed with the help of image processing techniques. Hence, the resolution between millimeter and pixel is needed with high accuracy. In cephalometric analysis, a scale is placed in front of patient's head when taking the radiograph. This study aims to develop an algorithm to recognize the scale in cephalogram, locate the scale mark and calculate the pixel-millimeter-ratio. First, a ROI is detected and cropped based on regression tree voting. Second, an algorithm is employed in ROI to detect the corner points of the scale and rotate the scale to perfectly vertical direction. Finally, a pixel tracing algorithm is employed to locate the first and the last scale mark in order to calculate the pixel length of the calibration. A novel method is proposed to adaptively assign the size and orientation of the image patches described by the SIFT vectors to ensure invariance of scale and rotation of images. The algorithm is robust to interference including tags, stickers and stain. A dataset consisting 163 cephalograms is tested and the algorithm performs a 100% Success Detection Rate within the precision range of 1.0mm.

*Cephalogram, Scale detection, SIFT descriptor, regression tree, edge detection*

## I. Introduction (Heading 1)

Cephalometric analysis is a widely used tool for assessment and prediction of craniofacial growth, orthodontic diagnosis, and oral-maxillofacial treatment planning for patients with malocclusion in clinical practice [1]. The traditional manual cephalometric analysis was firstly introduced by Hofrath [2] in 1931. The manual processing is very complicated with poor accuracy and repeatability. At present, most cephalometric analysis is done with the help of computer and digital image processing. A computer-automated cephalometric analysis system has been developed, which consists of four main steps:

- Cephalometric landmark detection;
- Forehead landmark-detection in profile photograph;
- Registration between radiograph and photograph;
- Measuring angular and linear parameters using the landmark locations.

In the system, the first step is realized by a decision tree regression voting method using SIFT based patch features [3]. The third step is realized by a novel contour-based registration method [4]. The mentioned algorithms require the cephalograms to be in the same resolution, scale and pixel-mm ratio. However, in clinical practice, the image set obtained varies in many aspects and may be in poor image quality with different rotations as well. In the system, pixel distance of landmarks is measured which should be translated to true distance in millimeter. Therefore, the transformation between true distance and pixel distance is needed. In orthodontic analysis, a Calibrated scale with a specific length is placed in front of patient's head as distance indicator when taking radiograph as illustrated in Fig.1(a).

Although, there are some baseline scale mark detection algorithms in other field [5], the implements require the scale to be full in picture view and a relatively fixed position. The practical circumstance, relatively unconstrained scale position and interferences include tags in Fig.1(b) and stickers in Fig.1(c), demands a more robust algorithm.

This study aims to develop an automatic method to recognize the calibrated scale in cephalogram and calculate the pixel-millimeter-ratio. First, a ROI (region of interest) extraction method based on SIFT (scale invariant feature transform) descriptor and regression tree voting is employed to find a region that coarsely contains the scale. The extraction method is similar to the landmark detection method in [3] with less hierarchy and accuracy. Second, a multilevel contour detection algorithm is applied to ROI to detect the left edge of the scale. Third, the corner points of the scale are located on left edge of scale basing on SIFT descriptor comparison. In this section, a novel method is proposed to adaptively assign the size and orientation of the image patches according to prior coarse detection so that the extracted SIFT features are invariant to scale and rotation of images. Finally, the scale mark is traced to locate the uppermost and lowermost scale mark in order to detect the measurement length of the scale. The flow chart of the proposed algorithm is illustrated in Fig.2.

*Huiqi Li is the corresponding author (e-mail:huiqili@bit.edu.cn).

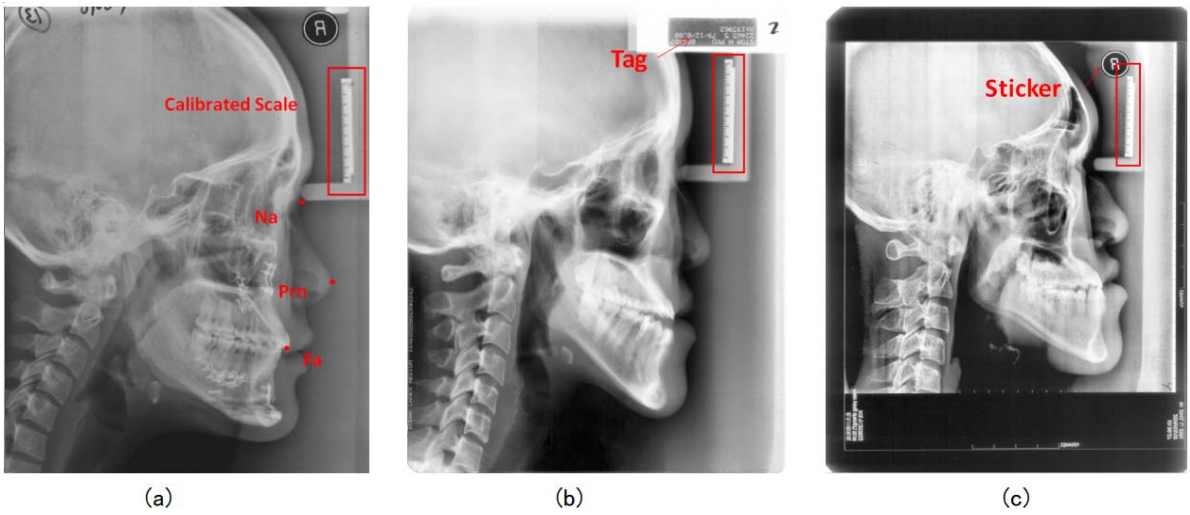

Figure 1. Calibrated scale and landmarks in cephalogram

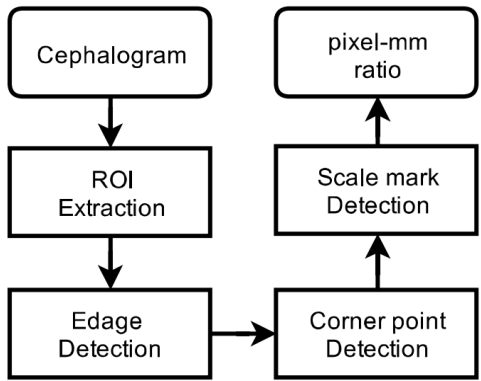

Figure 2. Flow chart of the proposed algorithm

II. METHOD

A. *ROI Extraction Method Based on SIFT Patch Descriptor and Regression Tree Voting*

A high-performance cephalometric landmark detection method is proposed in [3]. This method has a great accuracy and robustness on cephalogram for a main reason: the final location of the landmarks is decided by multiresolution voting of features of random patches in the whole image. Taking the advantage of most of the features of whole image, the final location predicted by regression tree will not deviate too much. As we want to extract ROI, the most important element to consider is stability. In this study, we employed a binary tree CART (classification and regression trees) [6] as the regressor to learn and predict the mapping relationship between the SIFT-based patch feature vectors $V$ and the displacement vectors $d(dx, dy)$ from the centers of the random patches to the position of scale corner points in the training images.

*1) SIFT Feature Descriptor.*

SIFT descriptor is used to represent local feature for corner points in the field of image matching, which is developed by Lowe in 2004[7]. The algorithm of SIFT feature extraction includes four steps:

- Scale-space extrema detection: potential key points are extracted in scale-space by using difference-of-Gaussian image pyramids;
- Key point localization: determine the location and scale of the key point;
- Orientation assignment: the orientation is assigned to the key point to provide invariance to rotation; and
- Keypoint descriptor: the local gradient is calculated in the selected scale which will provide invariance to shape distortion.

The advantage of SIFT features to represent the key points is affine invariant and robust to illumination change for images. This feature extraction method has been widely used in the fields of image matching and registration.

In this paper, we use the SIFT feature descriptors to represent image patches. An image patch centered at the selected key point with width of $2W+1$ pixels is divided into 4×4 small adjacent regions. The image gradient magnitudes and orientations of each pixel are calculated in the patch. Each region is described by 8-bin histogram of gradients representing the gradient magnitude of 8 main directions angles. In an image patch, 4×4 histograms of gradients are connected as feature vector $V$ (dimension=128).

*2) Training and Prediction of Scale Corner Points Based on Regression Tree*

Regression tree is a classical statistical learning algorithm used in regression problems. It learns the observation feature of an item to conclude the target value of the item. In this study, we employed CART (classification and regression trees) a binary tree as the regressor to learn the mapping relationship between the SIFT vectors $V$ of random selected image patches and the displacement vectors $d(dx, dy)$ from the centers of the patches to the location of upper left and lower left scale corner points in the training images. In the prediction, the regressors predict the displacements $d$ of a number of random selected patches and indicate the possible position of the scale corners. We firstly introduce four scales in the training and prediction of regression tree: 0.125, 0.25, 0.5, and 1 with a gradually increasing resolution. In the scale of 0.125, the patches are randomly selected in the whole image, which means the learning and prediction of patches is global. In the scale of 0.25, 0.5 and 1, the patches are randomly selected in the neighbourhood of the predicted location of the lower scale. Finally, the location of the scale corner point is obtained by the voting of the predicted displacements. The predicted corner point location is calculated as:

$$L = \frac{\sum_{i=1}^{m} p_i + d_i}{m} \quad (1)$$

where $p_i$, $d_i$ and $m$ is the location of the $i$th selected patch, the predicted displacement and the total number of random image patches. The prediction in 0.125-scale is illustrated in Fig.3(a) where the green line, blue point and red cross represent displacement $d$, $p_i + d_i$ and $L$ respectively. The four-scale regression tree prediction method is firstly used to predict the location of scale mark and the performance is tested with a dataset (see section 3). However, the accuracy and the Success Detection Rate are relatively low.

In this study, the accuracy is not the first priority in ROI extraction step. Hence, prediction in only two scales, 0.125 and 0.25 is employed to save the computational time. An example of prediction result is shown in Fig.3(b). The corner points are predicted and shown by red crosses.

*3) ROI Extraction and Preprocessing*

A ROI is cropped according to the location of predicted scale corner and the region is large enough to include the scale. An instance of ROI extraction is shown in Fig.3(c). To eliminate irregular illumination, pre-processing including contrast enhancement and Gamma correction is performed in the ROI, which is demonstrated in Fig.3(d).

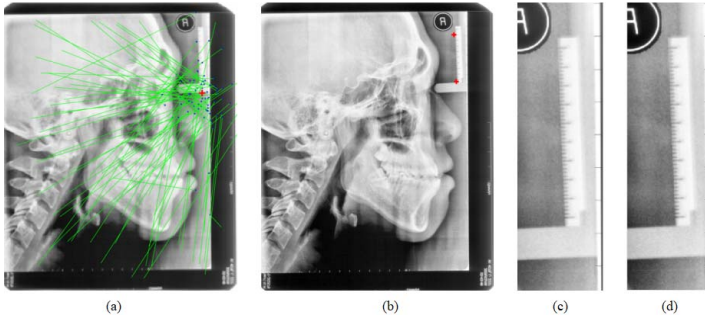

Figure 3. (a) Regression tree voting in scale 0.125. (b) Corner points prediction. (c) ROI extraction. (d) ROI preprocessing

*B. Scale Edge Detection*

The scale is long-edge-vertical with a slight incline. In the ROI, the scale has a higher greyscale comparing to the background. Therefore, we employ horizontal-response Sobel gradient operator $S$ to make convolution with the ROI; then, binarize the gradient image.

Before the edge detection, grayscale corrosion is applied to ROI to eliminate the influence of calibrated line illustrated in Fig.4(a). We define $A$ as the source image, and $G_x$ is the gradient image containing the horizontal derivative approximation. Operator $S$ and gradient image $G_x$ is represented as follow:

$$S = \begin{bmatrix} -1 & 0 & 1 \\ -2 & 0 & 2 \\ -1 & 0 & 1 \end{bmatrix} \quad (2)$$

$$G_x = S \times A \quad (3)$$

The positive result of $G_x$ representing an increasing greyscale as $x$ increase. Therefore, the positive high responses of $G_x$ appear at the left edge of the scale. An example of $G_x$ is illustrated in Fig.4(b).

Sometimes, there are influencing obstacles including tags (top left corner in Fig.4(a)), stickers and forehead in ROI. Besides, due to inhomogeneous illumination, the greyscale intensity of the scale may drift from the upper part to the lower part. These combined interferences will cause wrong or fragmentary recognition. In order to make sure the long edge of scale is detected, an adaptive binarization threshold $th_0$ is applied to make sure the edge detection result is long enough. In binarization procedure, the pixels in $G_x$ with intensity higher than $th_0$ will be set to one in binary image. We assume the approximate length of the scale as the pixel distance $L_p$ between the two corner points predicted in section 2.1. The longest connected domain (in $y$ coordinate) in binary image is denoted as $L_b$. We set the initial $th_0$ half of the highest intensity in $G_x$. The binarization will be repeated to decrease $th_0$ until $L_b > 0.6 \times L_p$. The binary image result of the first round and the final round are illustrated by Fig.4(c) and Fig.4(d) respectively.

Finally, linear Hough transformation [8] is applied to the binary image to locate the line of edge and calculate the inclination. The linear Hough transform algorithm uses a two-dimensional Hough space $(r, \theta)$ to detect the location of a line described as:

$$r = x \cdot cos\theta + y \cdot sin\theta. \quad (4)$$

For each pixel at $(x, y)$ and its neighborhood, the Hough transform algorithm determines if there is enough evidence of a straight line at that pixel. If so, it will calculate the parameters $(r, \theta)$ of that line, and then look for the Hough space's point that the parameters fall into, and increment the value of that point. By looking for local maximum in the Hough space $(r, \theta)$, the most likely lines can be extracted. The local maximum represents the long line in the image. Occasionally, the frame of the cephalogram can be detected as long line. Therefore, the leftmost Hough line in ROI is picked as the edge of the scale. The result of Hough transformation is illustrated by Fig.4(e) in green line.

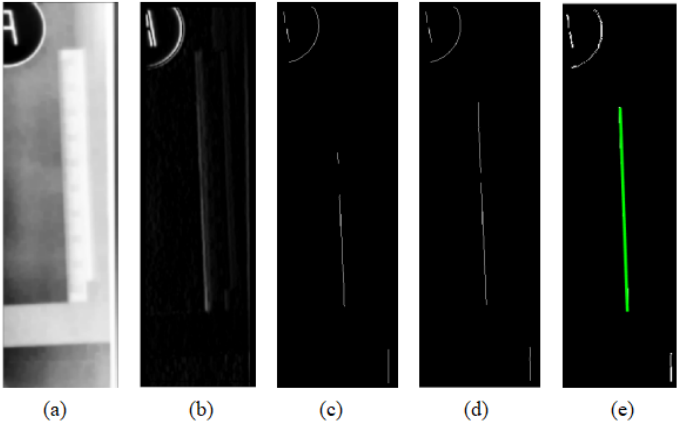

Figure 4. Scale edge detection on ROI. (a) Grey-scale Corrosion of ROI. (b) Horizontal-response Sobel edge detection. (c) Fragmentary edge binarization. (d) Complete edge binarization. (e) Hough transform.

*C. Corner Point Detection*

In this section, we use SIFT feature vector comparison to accurately locate the upper left and lower left corner point of the scale. A novel method is proposed to adaptively assign the size and orientation of the image patches according to coarse detection for the invariant SIFT feature extraction.

As discussed in section 2.1.1, an image patch, spanning a square window with width of $2W+1$ pixels, centered at location $(x, y)$ will be described by a 128-dimension vector $V$. In two images, two SIFT vectors $V$ that span the identical area of the same object is theoretically the same. Therefore, we compare between the SIFT descriptors $V_t$ centered near the left edge of scale in test ROI and two SIFT descriptors $V_r$ centered at two chosen corner points in reference ROI in order to pick out the point with least squared Euclidean distance as the corner point.

In traditional SIFT feature extraction, orientation is assigned in according to the main gradient angle for invariance to image rotation. For invariance to object scale, the size of patch is assigned according to the scale-space in the difference-of-Gaussian image pyramids. In our algorithm, since the image patches are selected according to location of edge detection, the scale and the orientation of patches that $V$ describe is assigned under the following rules:

- The main orientation of the image patch described by $V_t$ points along the edge of the scale in non-rotated ROI. The inclination angle of the edge is calculated by Hough transform in previous section. In another form, the image patches are assigned with orientations pointing vertically downward in rotated ROI where the scale edge is perfectly vertical. The ROI can be rotated according to the angle of Hough line resulted in section 2.2 as illustrated in Fig.5(b).

- The $V_r$ centered at chosen corner points in reference cephalogram span a square window of length $2W + 1$ where $W$ is chosen to be the width of the scale in reference ROI. Hence, $V_t$ should also have a $W$ equal to the width of the scale in test ROI to ensure scale invariance. In cephalogram, the width of the scale is in proportion to length, so we can estimate the width of scale in test ROI in accordance to the corner point distance $L_p$ predicted in section 2.1.

The $V_r$ centered at chosen corner points in reference ROI is illustrated in Fig.5(a). The searching areas of $V_t$ for upper corner and lower corner are selected near the edge and depicted by color strips as illustrated in Fig.5(c) and Fig.5(d) respectively. The red in the color strip represents less Euclidean distance between $V_r$ and $V_t$. The $V_t$ with least squared Euclidean distance to $V_r$ is chosen as illustrated in Fig.5(b). The pixel distance between the two corner points is denoted as $L_g$.

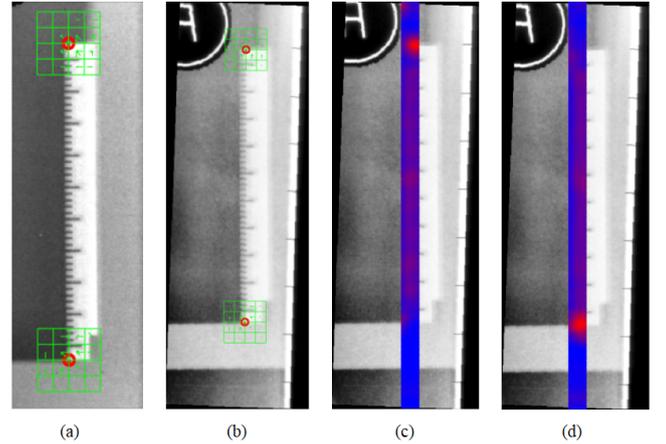

Figure 5. (a) $V_r$ in reference ROI. (b) $V_t$ in test ROI. (c) Searching area of upper corner. (d) Searching area of lower corner.

*D. Scale Mark Detection*

The calibration length is defined as the distance between the uppermost scale mark and the lowermost scale mark. In the cephalogram data set tested in this study, the length of scale is 45mm. On the integer multiple of 5mm, the major scale mark is longer than the rest. In this section, we trace a pixel column which is perfectly parallel to the long edge of scale from the lower corner point as a starting point; each gray-scale local minimum in the pixel column is a scale mark. In some cephalograms, the short scale marks are blurry; hence, a pixel column that only passing major scale mark is determined by algorithm. Occasionally, there are stains on the scale. A period-detection algorithm is employed to eliminate the influence of unexpected local minimum.

An algorithm is proposed to locate long scale mark and measure the distance which includes four steps as follow.

*1) Scale Mark Tracing*

We set an offset variable $off_x$, initially to 10. From the coordinate of the lower corner point with x coordinate add by $off_x$, we trace a column line $Line_c$ with length $L_g + 20$ which is properly longer than the distance of two corner point predicted in section 2.3. The line is illustrated in Fig.6(a) and the gray-scale of the line is shown in Fig.6(b). A black top-hat transformation is employed to this pixel line in order to eliminate inhomogeneous illumination and turn the scale mark to gray-scale maximum. The black top-hat transform is defined as the difference between the closing image and the input image:

$$T_b = L \cdot b - L \quad (5)$$

where · is the closing operation, $b$ is the structuring element and $L$ is the original line, i.e. $Line_c$. The closing operation and top-hat transformation result is shown in Fig.6(c) and Fig.6(d) respectively. The local maximums that exceed a threshold $th_1$ are recognized as scale mark.

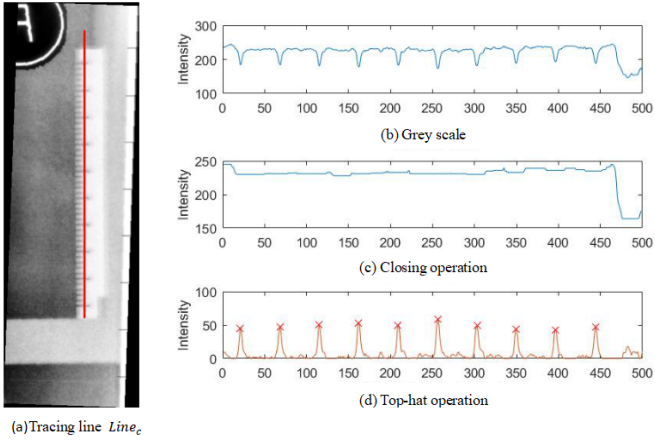

Figure 6. (a) Tracing line $Line_c$. (b) Gray-scale intensity of $Line_c$. (c) Closing operation of $Line_c$. (d) Black top-hat transform of $Line_c$.

*2) Major Scale Mark Judging*

In order to ensure the pixel line $Line_c$ passing only major scale mark, the distances of two adjacent local maximums is measured. If the average distance is shorter than $0.05L_g$, the pixel line is passing through short scale mark; then, step one (scale mark tracing) and two (Major scale mark judging) is repeated with $off_x$ increased by 3. If not, the pixel line only passes through long scale mark which makes it easy to locate each long scale mark.

*3) Stain Elimination*

Occasionally, there are stains on the scale in some cephalograms as shown in Fig.7(a) and Fig.7(b). The stain may be recognized as a local maximum in $Line_c$ which may confuse the scale mark. In order to eliminate the influence of the stain, the pixel distance interval between all adjacent local maximums are calculated and saved in an array: Local Maximum Interval (LMI). If there is no stain, all elements in LMI will be approximately the same, which is the period the scale mark appears on $Line_c$. If there are stains, the intervals between two adjacent scale marks will be segmented due to unexpected local maximum. In the algorithm, the small interval elements in LMI will merge to an interval that is big enough to fall around the biggest interval in LMI in order to filter the unexpected local maximum.

For instance, the top-hat $Line_c$ of Fig.7(a) is illustrated in Fig.7(c). The LMI before filtered and the LMI after filtered is [55 14 7 34 54 56 56 56 55 55] and [55 55 54 56 56 56 55 55] respectively. As a result, the third and fourth local maximum can be eliminated.

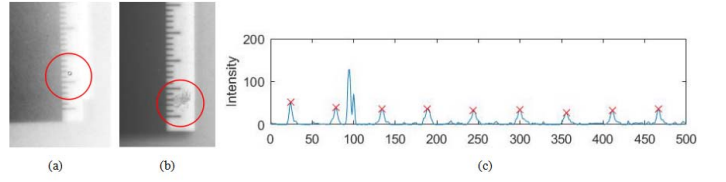

Figure 7. (a) Stains on scale. (b) Stains on scale. (c) Stain elimination.

*4) Calibration Length Calculation*

The scale marks can be located on the rotated ROI by detecting local maximum on top-hat $Line_c$. Then the locations are mapped to the original ROI without rotation as shown in Fig.8(a). The calibration length, i.e. distance between the first and the tenth major scale mark, is measured. The scale marks can also be mapped to the whole cephalogram as shown in Fig.8(b). The pixel-millimeter-ratio(PMR) can be calculated as 45mm divided by the pixel distance between the first and the tenth scale mark:

$$\text{PMR} = \frac{45}{|loc1 - loc10|} \quad (6)$$

where loc1 and loc10 represent the location of the first and the tenth scale mark respectively.

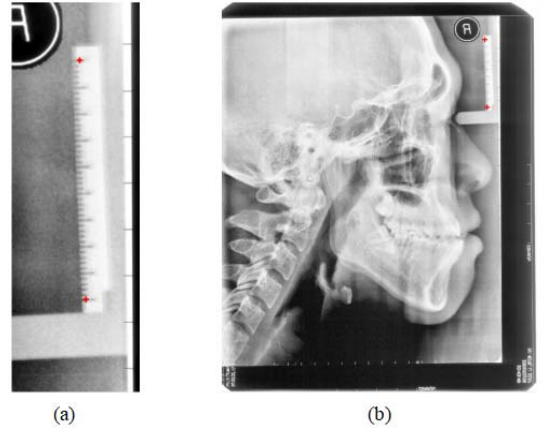

Figure 8. (a) Scale marks location on non-rotated ROI. (b) Scale marks location on cephalogram.

### III. EXPERIMENTAL RESULTS

#### A. Data Set

The dataset was obtained from the Peking University Hospital of Stomatology. The samples consist of 163 Chinese young adult subjects (the IRB approval number is PKUSSIRB-201415063). The image resolution varies from $758 \times 925$ to $2690 \times 3630$ pixels. Before processing, all images are normalized to the fixed width of 1960 pixels in dataset. In this dataset, the calibration length is 45mm. The uppermost and lowermost scale marks were manually annotated in each cephalogram for evaluation purpose and the pixel-millimeter-ratio is calculated.

## B. Evaluation

There are two evaluation criterions adopted in the result analysis. The first criterion includes the Mean Real Length Error(MRE) and Mean Pixel Length Error(MPE). The Pixel Distance Error $Ep$ is defined by the pixel difference between the predicted calibration length and the manually annotated calibration length, which is illustrated as:

$$Ep = |lp - lm| \quad (7)$$

where $lp$ and $lm$ is the predicted calibration length using the proposed algorithm and manually annotated calibration length respectively. The Real Length Error $Er$ is defined by the multiplication between $Ep$ and millimeter-to-pixel ratio of the manually annotated scale, which is illustrated as:

$$Er = Ep \times \frac{45}{lm} \quad (8)$$

The Mean Real Length Error (MRE) is the mean value of $Er$ in test set. The Mean Pixel Length Error (MPE) is the mean value of $Ep$ in test set.

The second evaluation criterion is the success detection rate (SDR) with respect to the 0.25 mm, 0.5 mm and 1 mm precision ranges. If $Er$ is less than a precision range, the detection of the scale is considered as a successful detection in the precision range; otherwise, it is considered as a failed detection. The SDR with Real Length Error less than error precision p is defined as:

$$SDR = \frac{\{Er_i < p\}}{M} \times 100\%, 1 \leq i \leq M \quad (9)$$

where $M$ is the total number of cephalograms in the test set and $p$ denotes three precision range including 0.25mm, 0.5mm and 1mm.

## C. Result and Discussion

The experimental results of both our proposed algorithm and four-scale regression tree prediction (4-rtp) are assessed and shown in Table1. In our algorithm, the MPE and MRE are 1.74 pixel and 0.17mm respectively. The SDRs are 75.31%, 96.91% and 100% within the precision ranges of 0.25mm, 0.5mm and 1mm respectively. The result is much better than the result of 4-trp.

TABLE I. EXPERIMENTAL RESULTS

|       | MPE (pixel) | MRE (mm) | SDR(%) 0.25mm | SDR(%) 0.5mm | SDR(%) 1mm |
|-------|-------------|----------|---------------|--------------|------------|
| Ours  | 1.74        | 0.172    | 75.31%        | 96.91%       | 100%       |
| 4-rtp | 4.5         | 0.458    | 47.31%        | 75.27%       | 92.47%     |

The MRE is 0.17mm which means that the average error comparing to manual annotation is far less than 1mm. The SDR with respect to precision 1mm is 100% which indicates that the Real distance Error will not exceed 1mm in all cephalograms in test set. The quantitative assessment indicates the high robustness and accuracy of the proposed algorithm. Some cephalograms in this dataset are attached with paper labels, tags and stains before digitalization which may cause the failure of scale detection as discussed in section 2.2 and 2.3.3. The quantitative assessment proves that the proposed algorithm is very robust to the above mentioned interference.

## IV. CONCLUSION

In conclusion, we proposed an algorithm of calibrated scale recognition in lateral cephalogram and the pixel-millimeter-ratio is calculated which is important in automatic cephalometric analysis. The scale can be accurately detected and located by the algorithm that includes ROI cropping, edge detection and corner point detection. The major scale mark can be located by tracing a pixel line. In addition, a novel method is proposed to assign the scale and orientation of image patches for the invariant SIFT feature extraction. The image patches chosen without difference-of-Gaussian image pyramids can be assigned size and orientation according to the estimated size and rotation of the object related to the image patch.

The algorithm can work well even under the interfering condition of tags, stickers and stains owing to a multilevel contour detection algorithm and a stain filter method. As demonstrated in experimental result, the algorithm has obtained satisfactory accuracy and robustness. In the test set, the algorithm performs a 100% Success Detection Rate within precision range of 1.0mm.

Although the algorithm is designed for a certain pattern of calibrated scale, it can be easily modified in accordance to the pattern of scale in other datasets of cephalograms. The algorithm can be further extended to other application of scale detection such as water level scale.